\documentclass[epj]{svjour}
\usepackage{graphics}
\begin{document}
\title{ Phase transition of compartmentalized surface models}

\author{Hiroshi Koibuchi
}                     
%
%
\institute{Department of Mechanical and Systems Engineering \\
  Ibaraki National College of Technology \\
  Nakane 866, Hitachinaka, Ibaraki 312-8508, Japan }
%
%
\abstract{
Two types of surface models have been investigated by Monte Carlo simulations on triangulated spheres with compartmentalized domains. Both models are found to undergo a first-order collapsing transition and a first-order surface fluctuation transition. The first model is a fluid surface one. The vertices can freely diffuse only inside the compartments, and they are prohibited from the free diffusion over the surface due to the domain boundaries. The second is a skeleton model. The surface shape of the skeleton model is maintained only by the domain boundaries, which are linear chains with rigid junctions. Therefore, we can conclude that the first-order transitions occur independent of whether the shape of surface is mechanically maintained by the skeleton (= the domain boundary) or by the surface itself.
}
\PACS{
      {64.60.-i}{General studies of phase transitions} \and
      {68.60.-p}{Physical properties of thin films, nonelectronic} \and
      {87.16.Dg}{Membranes, bilayers, and vesicles}
} 
\authorrunning {H.Koibuchi}
\titlerunning {Phase transition of compartmentalized surface models}
\maketitle
%
\section{Introduction}\label{intro}
The crumpling transition of membranes is an interesting topic in the softmatter physics as well as in the biological physics \cite{NELSON-SMMS2004,Gompper-Schick-PTC-1994,Bowick-PREP2001}. A well-known model for such transition is the surface model of Helfrich, Polyakov, and Kleinert \cite{HELFRICH-1973,POLYAKOV-NPB1986,KLEINERT-PLB1986}. A considerable number of theoretical and numerical studies have been devoted to reveal the phase structure of the model \cite{Peliti-Leibler-PRL1985,DavidGuitter-EPL1988,PKN-PRL1988,KANTOR-NELSON-PRA1987,KD-PRE2002,KOIB-PRE-2005,KOIB-NPB-2005,Baum-Ho-PRA1990,CATTERALL-NPBSUP1991,AMBJORN-NPB1993,KOIB-EPJB-2005}. Recently, it was shown by Monte Carlo (MC) simulations that the model undergoes a first-order transition on spherical and fixed connectivity surfaces \cite{KD-PRE2002,KOIB-PRE-2005}, and the transition is universal \cite{KOIB-NPB-2005}. The vertices can move only locally on the surface because of the fixed connectivity nature in those surface models.

However, the crumpling transition is not yet clearly understood in biological membranes. If we consider the possibility of the transition in the cell membranes, we should take account of the fluid nature such as the lateral diffusion of lipids. 

Conventionally, the free diffusion of lipids has been realized by the dynamical triangulation technique in the surface model \cite{Baum-Ho-PRA1990,CATTERALL-NPBSUP1991,AMBJORN-NPB1993,KOIB-EPJB-2005}. The diffusion of lipids has no cost in energy on fluid surfaces. 

In the cell membranes, however, the free diffusion of membrane proteins and lipids is suffered from heterogeneous structures. Such molecules are known to undergo the so-called hop diffusion over the surface, which was recently found experimentally \cite{Kusumi-BioJ-2004}. The free diffusion of the molecules is prohibited due to the cytoskeleton. The diffusion rate is, therefore, 10-100 times lower than that of artificial membranes, which are usually homogeneous and have no such domain structure. Moreover, some artificial membranes are considered to have skeletons, because they are partly polymerized \cite{CNE-PRL-2006}. 

Motivated by this fact observed in the cell membranes, we study firstly in this paper a dynamically triangulated surface model with compartmentalized domains whose boundaries are composed of triangle edges (or bonds) that are not to be flipped. The diffusion is constrained so that vertices can diffuse only inside the compartments, and hence the vertices never jump across from one compartment to the other compartments. Nevertheless, we consider that such constrained lateral diffusion can simulate the hop diffusion in the cell membranes as the first approximation.  

It is also interesting to see whether the transition occurs in a simplified skeleton model, where only skeletons maintain the mechanical strength of the surface. Skeleton models for the cytoskeleton were already investigated in \cite{BBD-BioPJ-1998}. A hard-wall and hard-core potential was assumed on the polymer chains with junctions, and the responses to some external stress and the compression modulus were obtained \cite{BBD-BioPJ-1998}. Giant fluid vesicles coated with skeletons was experimentally investigated, and the mechanical properties were reported \cite{HHBRMC-PRL-2001}, where the actin filaments introduce an inhomogeneous structure in the homogeneous artificial membrane.

In \cite{KOIB-JSTP-2007-1}, the phase structure of a surface model with skeleton was investigated, and it was reported that the model has a first-order transition between the smooth phase and the crumpled phase. The interaction of the model in \cite{KOIB-JSTP-2007-1} is described by a one-dimensional bending energy for linear chains (or bonds) and the two-dimensional bending energy for junctions. The two-dimensional Gaussian bond potential is also assumed in the Hamiltonian. Consequently, a simplified skeleton model can be obtained from the model in \cite{KOIB-JSTP-2007-1} by replacing the elastic junction with a rigid junction. 

Therefore, we study secondly in this paper the rigid junction model by using the canonical Monte Carlo simulations and see how the transition of \cite{KOIB-JSTP-2007-1} occurs in such a simplified model. We must note that the rigid junction model is not identical to the elastic junction model in \cite{KOIB-JSTP-2007-1}. In fact, there are two types of elasticity at the junctions; one is the out-of-plane elasticity and the other is the in-plane elasticity. The former elasticity can be rigid in the limit of infinite bending rigidity $b_J\!\to\! \infty$ in the elastic junction model of \cite{KOIB-JSTP-2007-1}, however, the in-plane elasticity can not be controlled in the elastic junction model. Therefore, the rigid junction model in this paper and the elastic junction model in  \cite{KOIB-JSTP-2007-1} are considered to be two different models. 

\section{Fluid Surface Model}\label{fluid-model}
By dividing every edge of the icosahedron into $\ell$ pieces of the uniform length, we have a triangulated surface of size $N\!=\!10\ell^2\!+\!2$ (= the total number of vertices). The starting configurations are thus characterized by $N_5\!=\!12$ and $N_6\!=\!N\!-\!12$, where $N_q$ is the total number of vertices with the co-ordination number $q$. 

The compartmentalized structure is built on the surface by keeping the boundary bonds unflipped in the MC simulations with dynamical triangulation. The boundary of the compartment is constructed from a sequence of bonds that remain unflipped. The total number $N_C$ of compartments depends on the surface size $N$. We fix $n$ the total number of vertices inside a compartment to the following values:
\begin{eqnarray}
\label{number-inside}
&&n=21,\;36,\;66, \;91,\;120,\\
 &&(\# \;{\rm of\; vertices\; in \;a \;compartment}). \nonumber
\end{eqnarray}
As a consequence, $N_C$ is increased with the increasing $N$. The reason why we fix $n$ is that the size of compartment is considered to be finite, and then it is expected that total number of lipids in the compartment also remains finite in the cell membranes. We must emphasize that the finiteness of $n$, rather than the value of $n$, is physically meaningful, because we do not always have one to one correspondence between the vertices and the lipid molecules.  

\begin{figure}[htb]
\unitlength 0.1in
\begin{picture}( 0,0)(  0,0)
\put(6.5,-1.0){\makebox(0,0){(a) $(N,n)\!=\!(2562,21)$ }}%
\put(24,-1.0){\makebox(0,0){(b) $(N,n)\!=\!(15212,66)$ }}%
\end{picture}%
\centering
\resizebox{0.49\textwidth}{!}{%
\includegraphics{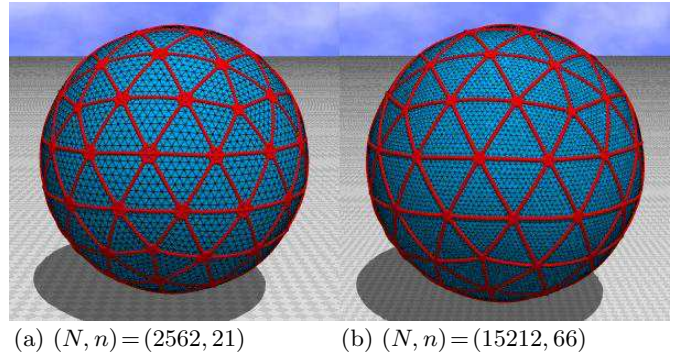}
}
\vspace{0.5cm}
\caption{Starting configuration of surfaces of (a) $(N,n)\!=\!(2562,21)$ and  (b) $(N,n)\!=\!(15212,66)$, where $n$ is the total number of vertices inside one compartment. Thick lines denote the compartment boundary consisting of bonds.} 
\label{fig-1}
\end{figure}

Figures \ref{fig-1}(a),(b) show surfaces of $(N,n)\!=\!(2562,21)$ and $(N,n)\!=\!(15212,66)$  for the starting configurations of MC simulations. Thick lines denote the compartment boundary consisting of the bonds that are not to be flipped. Vertices on the boundary of compartments can locally fluctuate, and they are prohibited from the diffusion. The remaining vertices freely diffuse only inside the compartments.
The starting configurations in Figs. \ref{fig-1}(a),(b) for the fluid model simulations are almost identical to those for the skeleton model simulations, which will be defined in the following section.

We note that the fixed connectivity surface model is obtained from the compartmentalized fluid model in the limit $n\!\to\!1$, where the vertices are prohibited from the free diffusion. On the contrary, we obtain the fluid surface model in the limit of $n\!\to\!N$, where all the vertices freely diffuse over the surface. 

The compartmentalized fluid surface model is defined by the partition function 
\begin{eqnarray} 
\label{Part-Func-F}
 Z = \sum_{ T} ^\prime\int^\prime \prod _{i=1}^{N} d X_i \exp\left[-S(X,{ T})\right],\\  
 S(X,{ T})=S_1 + b S_2, \nonumber
\end{eqnarray} 
where $b$ is the bending rigidity, $\int^\prime$ denotes that the center of the surface is fixed in the integration. $S(X,{ T})$ denotes that the Hamiltonian $S$ depends on the position variables $X$ of the vertices and the triangulation ${ T}$. $\sum_{ T}^\prime$ denotes the sum over all possible triangulations ${ T}$, which keep the compartments unflipped. The Gaussian term $S_1$ and the bending energy term $S_2$ are defined by
\begin{equation}
\label{Disc-Eneg-F} 
S_1=\sum_{(ij)} \left(X_i-X_j\right)^2,\quad S_2=\sum_{(ij)} (1-{\bf n}_i \cdot {\bf n}_j),
\end{equation} 
where $\sum_{(ij)}$ in $S_1$ is the sum over bonds $(ij)$ connecting the vertices $i$ and $j$, and $\sum_{(ij)}$ in $S_2$ is also the sum over bonds $(ij)$, which are edges of the triangles $i$ and $j$. The symbol ${\bf n}_i$ in $S_2$ denotes the unit normal vector of the triangle $i$. We emphasize that the compartment boundary gives no mechanical strength to the surface in this model.  

The bending rigidity $b$ has unit of $kT$, where $k$ is the Boltzmann constant, and $T$ is the temperature. The surface tension coefficient $a$ of $S_1$ is fixed to be $a\!=\!1$; this is always possible because of the scale invariant property of the model. In fact, in the expression $aS_1 \!+\! b S_2 $ we immediately understand that $a\!=\!1$ is possible, because the coefficient $a$ of $S_1$ can be eliminated due to the scale invariance of the partition function. Since the unit of $a$ is $(1/{\rm length})^2$, the length unit of the model is given by $\sqrt{1/a}$. We use the unit of length provided by $\sqrt{1/a}\!=\!1$ in this paper, although $a$ is arbitrarily chosen to be fixed.  

\section{Skeleton Model}\label{skeleton-model}
The model is defined on a triangulated surface, which is characterized by $N$ the total number of vertices including the junctions, $N_S$ the total number of vertices on the chains, $N_J$ the total number of junctions, and $L$ the length of chains between junctions. The junctions are assumed as rigid plates; twelve of them are pentagon and the others are hexagon. It should be noted again that $N_J$ is included in $N$; a junction is counted as a vertex. 

The surface of size $(N,N_S,N_J,L)\!=\!(2322,600,42,6)$ corresponds to that shown in Fig.\ref{fig-1}(a) for the fluid model. The reason why $N\!=\!2322$ of the surface is smaller than $N\!=\!2562$ of the one in Fig.\ref{fig-1}(a) is because the junctions are rigid objects. One hexagonal junction reduces $N$ by $6$, and one pentagonal junction also reduces $N$ by $5$ if they were assumed as rigid objects. Thus, we can check that $2562\!=\! 2322\!+\!6\times 30 \!+\!5\times 12$, where $30$ and $12$ are the total number of pentagonal junctions and that of hexagonal junctions, respectively.

We fix the chain length $L$ such that
\begin{equation}
\label{chain-length}
L=6\;(n=21),\quad L=11\;(n=66),
\end{equation}
 which respectively correspond to the values $n\!=\!21$, $n\!=\!66$; the total number of vertices inside a compartment in Eq.(\ref{number-inside}). The reason why we fix $L$ is the same as that for $n$.

The Hamiltonian of the model is given by a linear combination of the two-dimensional Gaussian bond potential $S_1$ and the one-dimensional bending energy $S_2^{(1)}$, which are defined by
\begin{equation}
\label{Disc-Eneg-S} 
S_1=\sum_{(ij)} \left(X_i-X_j\right)^2,\quad S_2^{(1)}=\sum_{(ij)} \left[1-\cos \theta_{(ij)}\right]. 
\end{equation} 
In these expressions, $\sum_{(ij)}$ in $S_1$ denotes the sum over all the bonds $(ij)$ connecting the vertices $i$ and $j$, and $\sum_{(ij)}$ in $S_2^{(1)}$ denotes the sum over bonds $i$ and $j$, which contain not only bonds in the chains but also {\it virtual bonds} that connect the center and the corners of the rigid junctions. The symbol $\theta_{(ij)}$ in $S_2^{(1)}$ is the angle between the bonds $i$ and $j$, which include the virtual bonds. The Gaussian potential $S_1$ is defined on all the bonds including those on the chains. As a consequence, the model is considered to be a surface model, although the mechanical strength is maintained by one-dimensional elastic skeletons joined to each other at the rigid junctions.

\begin{figure}[htb]
\centering
\includegraphics{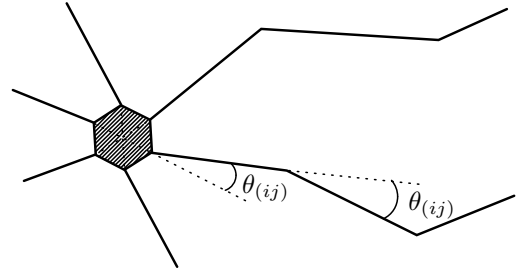}
\caption{A hexagonal junction connected to chains. The angle $\theta_{(ij)}$ in $S_2^{(1)}$ is defined not only at the vertices on the chains but also at the corners (=virtual vertices) of the junction. The triangular lattices attached to the chains were eliminated from the figure to clarify the chains. } 
\label{fig-2}
\end{figure}

Triangulated spherical surfaces for the skeleton model are almost identical to those shown in Figs.\ref{fig-1}(a),(b) as mentioned above. Only difference between the surfaces is whether the junctions are rigid objects or not. Figure \ref{fig-2} shows a hexagonal rigid junction connected to chains, where the angle $\theta_{(ij)}$ is defined not only at the vertices on the chains but also at the corners (={\it virtual vertices}) of the junction. The triangular lattices attached to the chains were eliminated from Fig. \ref{fig-2} to clarify the chains.

We should comment on the size of the junctions. The junctions are two-dimensional objects and therefore have their own size to be fixed. The size of junction can be specified by the edge length $R$; the perimeter length of the pentagonal (hexagonal) junction is therefore expressed by $5R$ ($6R$). In this paper, we fix the size of the junctions such that
\begin{equation} 
\label{junctionsize}
R=0.1.
\end{equation}
The value $R\!=\!0.1$ is quite smaller than that of the elastic junctions in \cite{KOIB-JSTP-2007-1}, where the edge length squared is $R^2\simeq 0.5$ because of the relation $S_1/N\!=\!1.5$, which is also satisfied in the fluid model defined in the previous section. 

We must note that the junction size in Fig. \ref{fig-2} were drawn larger, comparing to the bond length, than that expected from Eq.(\ref{junctionsize}). In the following section, we discuss how do we fix the size $R$ to the assumed value in Eq.(\ref{junctionsize}) in the MC simulations.

The partition function $Z$ of the skeleton model is defined by
\begin{equation} 
\label{Part-Func-S}
 Z = \int^\prime \prod _{i=1}^{N} d X_i \exp\left[-S(X)\right],\quad  
 S(X)=S_1 + b S_2^{(1)}, 
\end{equation} 
where $b$ is the bending rigidity corresponding to the one-dimensional bending energy, and  $\int^\prime$ denotes that the center of the surface is fixed. The integration $\prod _{i=1}^{N} d X_i$ is a product of the integration over vertices and that of junctions such that 
\begin{equation} 
\label{integration}
\prod _{i=1}^{N} d X_i = \prod _{{\rm vertices}\; i} d X_i  \prod _{{\rm junctions}\; i} d X_i, 
\end{equation} 
where $\prod _{{\rm junctions}\; i} d X_i$ is the integration over the degrees of freedom for the three-dimensional translations and rotations. 

\section{Monte Carlo technique}\label{MC-Techniques}
The vertices $X$ are shifted so that $X^\prime \!=\! X\!+\!\delta X$, where $\delta X$ is randomly chosen in a small sphere. The new position $X^\prime$ is accepted with the probability ${\rm Min}[1,\exp(-\Delta S)]$, where $\Delta S\!=\! S({\rm new})\!-\!S({\rm old})$. 

The summation over ${ T}$ in the fluid model partition function $Z$ of Eq.(\ref{Part-Func-F}) is performed by the standard bond flip technique. The bonds are labeled with sequential numbers. The total number of bonds $N_B$ is given by $N_B\!=\!3N\!-\!6$, which includes the bonds in the boundary of compartments.  Firstly, the odd-numbered bonds are sequentially chosen, to be flipped and secondly, the remaining even-numbered bonds are chosen. The flip is accepted with the probability ${\rm Min} [1, \exp(-\Delta S)]$. In this procedure, the compartment boundary remains unflipped. $N$ updates of $X$ and $N_B/2$ updates of ${ T}$ are consecutively performed and make one MCS (Monte Carlo Sweep). The radius of the small sphere for $\delta X$ is chosen so that the rate of acceptance for $X$ is about $50\%$, which is controlled by a small number for the radius $\delta X$ at the beginning of the simulations. We introduce the lower bound $1\times 10^{-8}$ for the area of triangles. No lower bound is imposed on the bond length.

The assumed sizes in the fluid model simulations are listed in Table \ref{table-1}. Three sizes of surfaces are assumed for each compartment size $n$ except for $n\!=\!120$. The third size for $n\!=\!91$ is relatively large and therefore time consuming for the fluid simulations, and the size $N\!=\!11562$ is sufficiently large to show that there is no phase transition on the surface of $n\!=\!120$.
\begin{table}[hbt]
\caption{The size $N$ of surfaces for the fluid model simulations. Three sizes of surfaces are assumed for each compartment size $n$ except for $n\!=\!120$. }
\label{table-1}
\begin{center}
 \begin{tabular}{ccccc}
$n\!=\!21$ & $n\!=\!36$ & $n\!=\!66$ & $n\!=\!91$ & $n\!=\!120$ \\
 \hline
  2562     & 1002       & 1692       & 2252       & 2892 \\
  5762     & 4002       & 6762       & 9002       & 11562 \\
  10242    & 9002       & 15212      & 20252      &  \\
 \hline
 \end{tabular} 
\end{center}
\end{table}

Total number of MCS after the thermalization MCS at $b$ close to the transition point is about $2\!\times\!10^8\sim 3\!\times\!10^8$ MCS on the $N\!\geq\! 9002$ surfaces and $1\!\times\!10^8\sim 1.7\!\times\!10^8$ MCS on the $N\!\leq\! 6762$ surfaces, and relatively smaller number of MCS at $b$ far from the transtion point. The thermalization MCS $1\!\times\!10^7\sim 1.7\!\times\!10^8$ MCS on the $N\!\geq\! 9002$ surfaces and $1\!\times\!10^7\sim 3\!\times\!10^7$ MCS on the $N\!\leq\! 6762$ surfaces. The reason of such a large number ($1.7\!\times\!10^8$) of thermalization MCS seems due to the discontinuous nature of the transition. The starting configurations of MC are just like those shown in Figs.\ref{fig-1}(a) and \ref{fig-1}(b), and hence they are in the smooth phase. In fact, large surfaces in the collapsed phase close to the transition point eventually collapsed after such a long thermalization MCS.

The update of $X$ in MC for the skeleton model partition function in Eq.(\ref{Part-Func-S}) can be divided into two steps, which are corresponding to the integrations $\prod _{{\rm vertices}\; i} d X_i$ and $\prod _{{\rm junctions}\; i}  d X_i$ in Eq.(\ref{integration}). The first is the update of $X$ of vertices including those in the chains. The second is the update of the position of the junctions as three-dimensional rigid objects. This can be further divided into two processes: the first is a random three-dimensional translation, and the second is a random three-dimensional rotation. All of these MC processes are independently performed under about $50\%$ acceptance rate.  

The junction size $R$ is fixed to $R\!=\!0.1$ in Eq.(\ref{junctionsize}) during the thermalization MCS. The initial value of $R$ is given by $R\simeq 0.7$ on the surfaces such as those shown in Figs.\ref{fig-1}(a) and \ref{fig-1}(b). Thus, we reduce $R$ from $R\simeq 0.7$ to $R\!=\!0.1$ by $\Delta R\!=\!6\!\times\! 10^{-6}$ at every $25$ MCS in the first $2.5\times 10^6$ MCS. Because of this forced reduction of the junction size, the equilibrium statistical mechanical condition seems to be violated in the first $2.5\times 10^6$ MCS. Therefore, relatively many thermalization ($1.75\times 10^7$ or more) MCS is performed after the first $2.5\times 10^6$ MCS for the reduction.

We use surfaces of size listed in Table \ref{table-2} for the skeleton model simulations. Four sized are assumed for $L\!=\!6$, and three sizes for $L\!=\!11$. 
\begin{table}[hbt]
\caption{The size $N$ of surfaces for the skeleton model simulations. Four sizes of surfaces are assumed for  $L\!=\!6$, and three sizes for $L\!=\!11$. }
\label{table-2}
\begin{center}
 \begin{tabular}{cccccccccc}
$L$ & $N$ & $N_S$ & $N_J$ & & $L$ & $N$ & $N_S$ & $N_J$ \\
 \hline
  6  & 5222   & 1350  & 92   &  &  11  & 6522   & 1200  & 42   & \\
  6  & 9282   & 2400  & 162  &  &  11  & 14672  & 2700  & 92   & \\
  6  & 14502  & 3750  & 252  &  &  11  & 26082  & 4800  & 162  & \\
  6  & 20882  & 5400  & 362  &  &      &        &       &      & \\
 \hline
 \end{tabular} 
\end{center}
\end{table}

The total number of MCS is about $1.6\!\times\!10^8\sim 2\!\times\!10^8$ for the $N\!\leq\!6522$ surfaces and $3\!\times\!10^8\sim 4\!\times\!10^8$ for the $N\!\geq\!9282$ surfaces. The thermalization MCS is $2\!\times\!10^7$ for the $N\!\leq\!6522$ surfaces and $2\!\times\!10^7\sim 1\!\times\!10^8$ for the $N\!\leq\!9282$ surfaces.

A random number sequence called Mersenne Twister \cite{Matsumoto-Nishimura-1998} is used in the simulations.
\section{Fluid Model Simulations}\label{Fluids}
\begin{figure}[htb]
\unitlength 0.1in
\begin{picture}( 0,0)(  0,0)
\put(7,33.5){\makebox(0,0){(a) $(20252,91),\; b\!=\!1.57$ }}%
\put(24,33.5){\makebox(0,0){(b) $(20252,91),\; b\!=\!1.58$ }}%
\put(6.5,-1.0){\makebox(0,0){(c) The surface section}}%
\put(24,-1.0){\makebox(0,0){(d) The surface section}}%
\end{picture}%
\centering
\resizebox{0.49\textwidth}{!}{%
\includegraphics{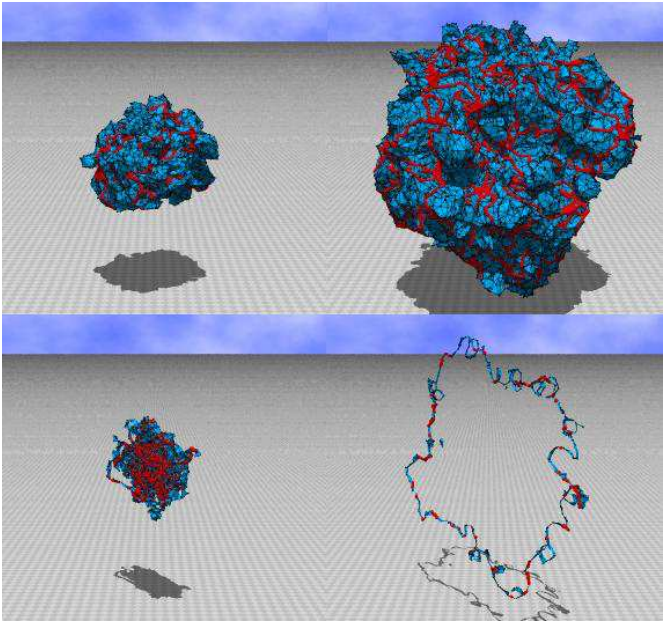}
}
\vspace{0.5cm}
\caption{Snapshot of the fluid surface of $(N,n)\!=\!(20252,91)$ obtained at (a) $b\!=\!1.57$ in the collapsed phase and at (b) $b\!=\!1.58$ in the smooth phase, and (c), (d) the surface sections. The mean square size $X^2$ defined in Eq.(\ref{X2}) is $X^2\!\simeq\!15$ in (a) and  $X^2\!-simeq\!151$ in (b).} 
\label{fig-3}
\end{figure}
First, we show in Figs. \ref{fig-3}(a),(b) snapshots of the $(N,n)\!=\!(20252,91)$ surfaces obtained at $b\!=\!1.57$ in the collapsed phase and at $b\!=\!1.58$ in the smooth phase.  Figures \ref{fig-3}(c),(d) show the surface sections. Thus, we confirmed that the smooth phase can be seen at finite $b$ close to the transition point.

\begin{figure}[htb]
\centering
\resizebox{0.49\textwidth}{!}{%
\includegraphics{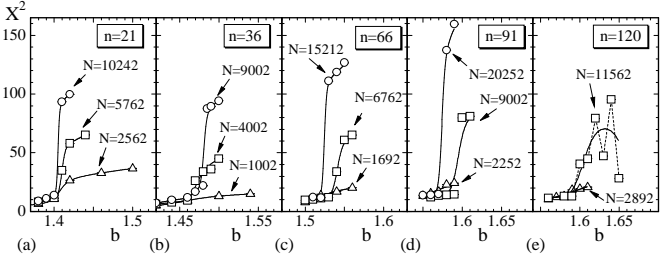}
}
\caption{The mean square size $X^2$ versus $b$ obtained on the fluid surfaces of (a) $n\!=\!21$, (b) $n\!=\!36$, (c) $n\!=\!66$, (d) $n\!=\!91$, and (e) $n\!=\!120$. Solid lines were obtained by the multihistogram reweighting technique.} 
\label{fig-4}
\end{figure}

The crumpling transition is conventionally understood as the one of surface fluctuation accompanied by surface collapsing phenomena, which can be seen in our surface model as we have seen in the snapshots in Figs.\ref{fig-3}(a)--\ref{fig-3}(d). Therefore, we expect that the mean square size $X^2$ reflects the collapsing transition on spherical surfaces. The mean square size $X^2$ is defined by
\begin{equation}
\label{X2}
X^2={1\over N} \sum_i \left(X_i-\bar X\right)^2, \quad \bar X={1\over N} \sum_i X_i,
\end{equation}
where $\bar X$ is the center of the surface. Figures \ref{fig-4}(a),(b),(c),(d), and (e) show $X^2$ obtained on the surfaces of $n\!=\!21$, $n\!=\!36$, $n\!=\!66$, $n\!=\!91$,  and $n\!=\!120$, respectively. Solid lines connecting the data were obtained by the multihistogram reweighting technique \cite{Janke-histogram-2002}. A discontinuous change of $X^2$ can be seen in the cases of $n\!\leq\! 91$ when the size $N$ increases, while $X^2$ largely fluctuates at $n\!=\!120$. This indicates that a first-order transition occurs at $n\!\leq\! 91$ and that it disappears at $n\!=\!120$. We also find that the transition point $b_n$ moves left on the $b$-axis as $n$ decreases. It is expected in the limit of $n\!\to\! 1$ that $b_n$ reduces to the value corresponding to the transition point of the fixed connectivity surface model \cite{KOIB-PRE-2005}. We can also confirm that $b_n$ disappears in the limit of $n\!\to\!N$ at sufficiently large $N$, because we find no transition at $n\!=\!120$.

\begin{figure}[htb]
\centering
\resizebox{0.49\textwidth}{!}{%
\includegraphics{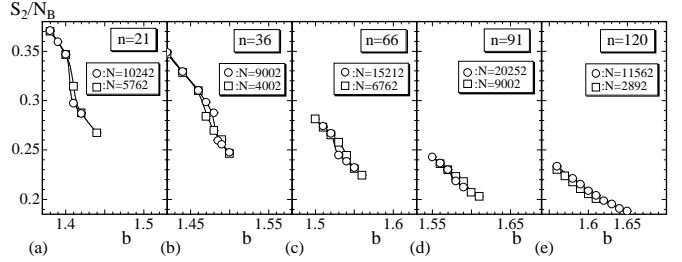}
}
\caption{The bending energy $S_2/N_B$ versus $b$ obtained on the surface of (a) $n\!=\!21$, (b) $n\!=\!36$, (c) $n\!=\!66$, (d) $n\!=\!91$, and (e) $n\!=\!120$. $N_B(\!=\!3N\!-\!6)$ is the total number of bonds.  } 
\label{fig-5}
\end{figure}
The crumpling transition is originally understood as the one for surface fluctuation phenomena. Therefore, the bending energy $S_2/N_B$, defined in Eq.(\ref{Disc-Eneg-F}), is expected to reflect the transition.  Figures \ref{fig-5}(a),(b),(c),(d) and (e) show $S_2/N_B$ versus $b$ corresponding to $n\!=\!21$, $n\!=\!36$, $n\!=\!66$, $n\!=\!91$, and $n\!=\!120$, respectively. Discontinuous change of $S_2/N_B$ can also be seen at $b$ where $X^2$ discontinuously changes, although it is not sufficiently clear in the figures.

\begin{figure}[htb]
\centering
\resizebox{0.4\textwidth}{!}{%
\includegraphics{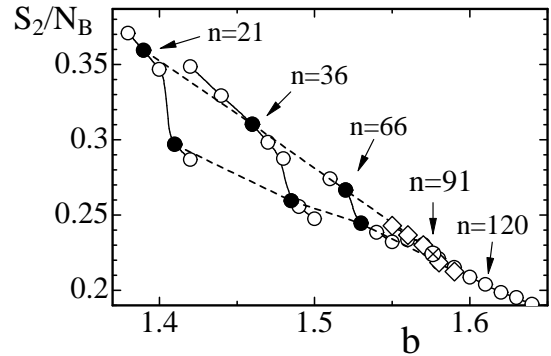}
}
\caption{The variation of the gap of $S_2/N_B$ against $n$, which were obtained on the surfaces of size  $(N,n)\!=\!(10242,21)$, $(9002,36)$, $(15212,66)$, $(20252,91)$, and $(11562,120)$.  The solid circles ($\bullet$) and the dashed lines denote the value of  $S_2/N_B$ in the smooth (the crumpled) phase at the first-order transition point. The diamonds ($\diamond $) correspond to the results obtained on the surface $(20252,91)$. The symbol ($\otimes $) denotes the critical point of transition, where the discontinuous transition terminates and turns to a continuous or a higher-order one. The critical value $n_c$ is expected to be $n_c\!\simeq\!91$. } 
\label{fig-6}
\end{figure}
In order to see the gap of $S_2/N_B$ more clearly, we show the variation of the gap of $S_2/N_B$ against $n$ in Fig.\ref{fig-6}. The solid circles ($\bullet$) and the dashed lines denote the value of  $S_2/N_B$ in the smooth (the crumpled) phase at the first-order transition point. The diamonds ($\diamond $) correspond to the results obtained on the surface $(20252,91)$. We find that the value of $S_2/N_B$ in the smooth phase (or in the crumpled phase) increases as $n$ decreases at the transition point. The value of $S_2/N_B$ at the transition point becomes identical to the one of the fixed connectivity surface model in the limit of $n\!\to\! 1$ \cite{KOIB-PRE-2005}. On the contrary, the discontinuous change of $S_2/N_B$ is expected to disappear at sufficiently large $n$, because the discontinuity of $S_2/N_B$ seen clearly at $n\!=\!21\sim 66$ disappears when $n\!\to \!N$ at sufficiently large $N$. This implies that there exists a finite $n_c$, where the first-order transition terminates and turns to a continuous or a higher-order one. The gap of $S_2/N_B$ at $n\!=\!21$ reduces as $n$ increases and eventually goes to zero at $n\!=\!n_c$, which is expected to be $n_c\!\simeq\!91$. At $n\!=\!91$, $S_2/N_B$ seems continuous while $X^2$ is discontinuous as confirmed in Fig.\ref{fig-4}(d). 

We note that the maximum co-ordination number $q_{\rm max}$ obtained throughout the simulations is as follows: $q_{\rm max}\!=\!38$ on the $(N,n)\!=\!(10242,21)$ surface at $b\!=\!1.38$, $q_{\rm max}\!=\!42$ on the $(N,n)\!=\!(9002,36)$ surface at $b\!=\!1.42$,  $q_{\rm max}\!=\!40$ on the $(N,n)\!=\!(15212,66)$ surface at $b\!=\!1.51$, $q_{\rm max}\!=\!42$ on the $(N,n)\!=\!(20252,91)$ surface at $b\!=\!1.55$, and $q_{\rm max}\!=\!44$ on the $(N,n)\!=\!(11562,120)$ surface at $b\!=\!1.56$.  

\begin{figure}[htb]
\centering
\resizebox{0.49\textwidth}{!}{%
\includegraphics{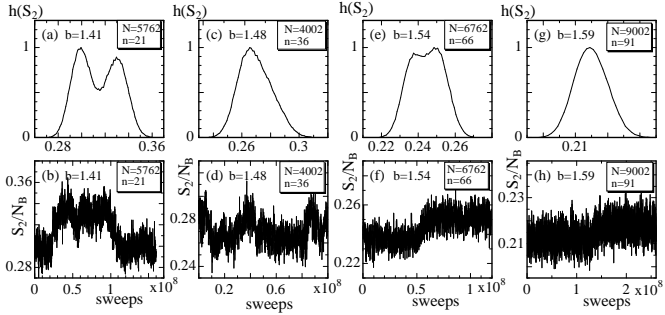}
}
\caption{The histogram $h(S_2)$ and the corresponding variation of $S_2/N_B$ against MCS obtained on the surfaces of (a),(b) $(N,n)\!=\!(5762,21)$,  (c),(d) $(N,n)\!=\!(4002,36)$, (e),(f) $(N,n)\!=\!(6762,66)$, and (g),(h) $(N,n)\!=\!(9002,91)$.   } 
\label{fig-7}
\end{figure}
In order to show the discontinuity in $S_2/N_B$ more clearly, we plot in Figs.\ref{fig-7}(a)--\ref{fig-7}(h) the distribution (or histogram) $h(S_2)$ of $S_2/N_B$ and the corresponding variation of $S_2/N_B$. These were obtained on the surfaces of $(N,n)\!=\!(5762,21)$,  $(4002,36)$, $(6762,66)$, and $(9002,91)$. The discontinuity of $S_2/N_B$ can be seen in the histogram on the $N\!\geq\!5762$ surfaces. Because of the size effect, the transition appears to be continuous on the $N\!=\!4002$ surface. The double peak at $(N,n)\!=\!(5762,21)$ is more clear than that at $(N,n)\!=\!(6762,66)$, because the gap of $S_2/N_B$ reduces as $n$ increases. We should note that the double peak structure is very hard to see in $h(S_2)$ on the $N\!\geq\!9002$ surfaces at $n\!=\!36$. When the configuration is once trapped in the smooth (the crumpled) state, it hardly changes to the crumpled (the smooth) state at the transition point on such large surfaces. This problem may be resolved with more sophisticated MC techniques \cite{Berg-Neuhaus-PRL1992,Berg-Celik-PRL1992}. No double-peak structure is found in Fig.\ref{fig-7}(g) and the variation of $S_2/N_B$ smoothly varies in Fig.\ref{fig-7}(h). This indicates that the transition is a continuous one at $n\!=\!91$ because of the fact that the surface of $(20252,91)$ becomes smooth at $b\!\geq\!1.58$ and crumpled at $b\!\leq\!1.57$, which was clarified from $X^2$ in Fig.\ref{fig-4}(d). Thus, the critical value was expected to be $n_c\!\simeq\!91$.

\begin{figure}[htb]
\centering
\resizebox{0.49\textwidth}{!}{%
\includegraphics{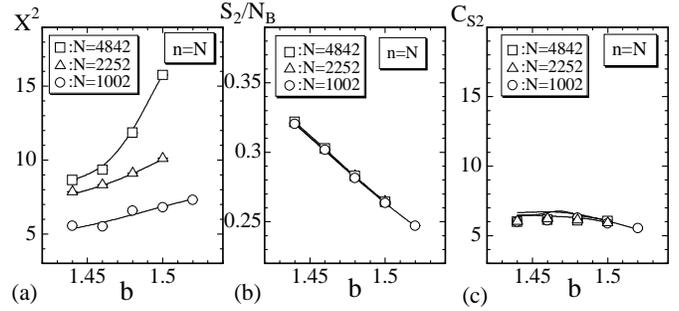}
}
\caption{(a) $X^2$ vs. $b$, (b) $S_2/N_B$ versus $b$, and (c) $C_{S_2}$ versus $b$ obtained on the fluid surface model without the compartmentalized structure. The symbol $n\!=\!N$ drawn on the figures denotes that the surfaces are those without the compartment.} 
\label{fig-8}
\end{figure}
The phase transition is expected to disappear from the fluid surface model defined on the surfaces that have no compartment. In order to show this, we performed MC simulations on the surfaces without the compartments up to the size $N\!=\!4842$. Figure \ref{fig-8}(a) shows $X^2$ versus $b$. We can see no abrupt growing of $X^2$ in the figure. The bending energy $S_2/N_B$ shown in Fig. \ref{fig-8}(b), where the variation of $S_2$ versus $b$ seems almost independent of $N$. The specific heat $C_{S_2}$, which is defined by 
$C_{S_2} \!=\! (b^2/ N) \langle \; \left( S_2 \!-\! \langle S_2 \rangle\right)^2\rangle$, 
is expected to reflect the phase transition. However, we can see no anomalous behavior in $C_{S_2}$ shown in Fig. \ref{fig-8}(c); there can be seen no peak in $C_{S_2}$. Thus we confirmed that the phase transition disappears from the model if $n\!\to \! N$ at sufficiently large $N$.

We must comment on the relation between the above results and those of fluid surface simulations in \cite{KOIB-PLA-2002}, because the phase transition can be seen in \cite{KOIB-PLA-2002} while it can not in Figs.\ref{fig-8}(a),(b). The difference between the bond flip procedure in this paper and that of \cite{KOIB-PLA-2002} is the reason why the phase transition can be seen in the model in \cite{KOIB-PLA-2002} and it can not be seen in the same model in Figs.\ref{fig-8}(a),(b). In the simulations of this paper we label the bonds by a sequence of numbers as stated above and perform the bond flip by using the sequential numbers. On the contrary, a vertex is firstly chosen randomly in \cite{KOIB-PLA-2002}, and secondly a bond is randomly chosen to be flipped from the bonds connecting the chosen vertex. As a consequence, this procedure gives a large (small) weight to the vertices which have small (large) co-ordination number in the dynamical triangulations. Therefore, the procedure in \cite{KOIB-PLA-2002} seems change effectively the coefficient $\alpha$ of the co-ordination dependent term $-\alpha \sum_i \log q_i$, which comes from the integration measure $\prod_i dX_i q_i^\alpha$, where $\alpha$ is fixed to $\alpha\!=\!0$ in this paper and in \cite{KOIB-PLA-2002}. We  know that the phase structure depends on the co-ordination dependent term in the fluid surface model \cite{KOIB-PLA-2003}.

\section{Skeleton Model Simulations}\label{Skeletons}
\begin{figure}[htb]
\centering
\vspace{0.7cm}
\unitlength 0.1in
\begin{picture}( 0,0)(  0,0)
\put(7.,31.5){\makebox(0,0){(a) $b\!=\!12.3$ (Collapsed) }}%
\put(23,31.5){\makebox(0,0){(b) $b\!=\!12.4$ (Smooth)}}%
\put(7,-1.0){\makebox(0,0){(c) The surface section }}%
\put(23,-1.0){\makebox(0,0){(d) The surface section}}%
\end{picture}%
\resizebox{0.49\textwidth}{!}{%
\includegraphics{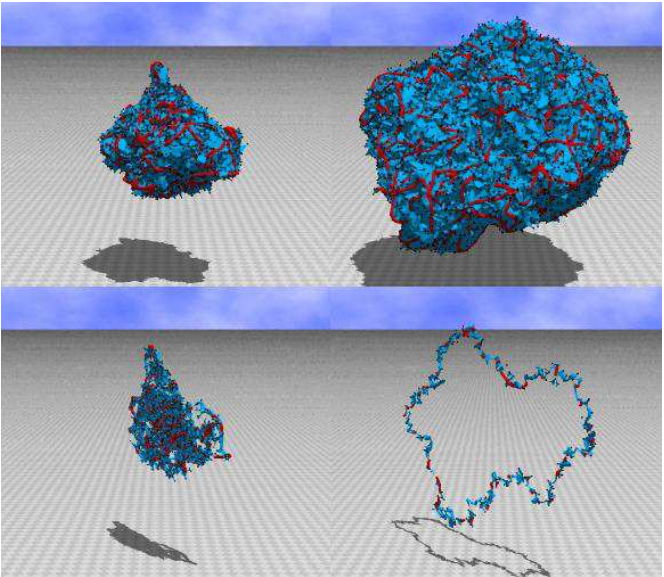}
}
\vspace{0.5cm}
\caption{Snapshot of the surface of size $(N,N_S,N_J,L)\!=\!(26082,4800,162,11)$ obtained in the collapsed phase at (a) $b\!=\!12.3$ and in the smooth phase at (b) $b\!=\!12.4$, both of which are close to the transition point. The mean square size $X^2$ is $X^2\!\simeq\!12$ in (a) and  $X^2\!\simeq\!98$ in (b). } 
\label{fig-9}
\end{figure}

Snapshots of the skeleton surfaces are shown in Figs.\ref{fig-9}(a)--\ref{fig-9}(d). Figure \ref{fig-9}(a) is a surface of size $(N,N_S,N_J,L)\!=\!(26072,4800,162,11)$ obtained at $b\!=\!12.3$ in the crumpled phase, and Fig.\ref{fig-9}(b) is the one obtained at $b\!=\!12.4$ in the smooth phase. The surface sections of Figs.\ref{fig-9}(a) and \ref{fig-9}(b) are shown in Figs.\ref{fig-9}(c) and \ref{fig-9}(d), respectively. These figures were drawn in the same scale. We immediately see the surface in the smooth phase is actually swollen, while the surface is collapsed in the collapsed phase. 

\begin{figure}[htb]
\centering
\resizebox{0.49\textwidth}{!}{%
\includegraphics{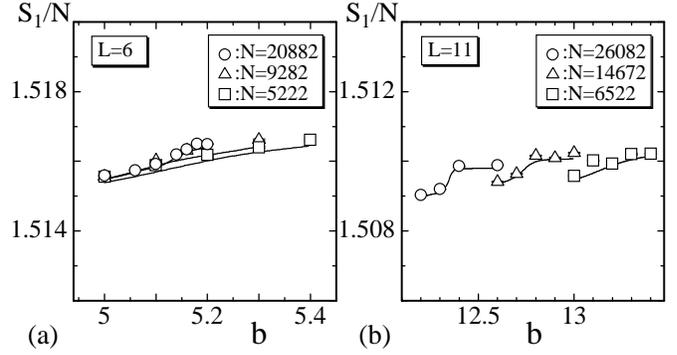}
}
\caption{The Gaussian bond potential $S_1/N$ versus $b$ obtained on the surfaces of (a) $L\!=\!6$ and (b) $L\!=\!11$.  $S_1/N$ slightly deviates from $S_1/N\!\simeq\!1.5$. } 
\label{fig-10}
\end{figure}
The Gaussian bond potential $S_1/N$ is shown against $b$ in Figs.\ref{fig-10}(a) and \ref{fig-10}(b), which correspond to the lengths $L\!=\!6$ and  $L\!=\!11$, respectively.
 The values of $S_1/N$ in the figures slightly deviate from $S_1/N\!=\!1.5$, which is satisfied on the surface without the rigid junctions or the rigid junctions of negligible size. The reason of this discrepancy is because the surface includes the rigid junctions of finite size. A vertex is the zero-dimensional point, while the rigid junction is a two-dimensional plate and hence shares an area of the surface. We find a gap or a jump in $S_1/N$ of the $(N,N_S,N_J,L)\!=\!(26082,4800,162,11)$ surface in Fig.\ref{fig-10}(b), which can be viewed as a sign of a discontinuous transition.

\begin{figure}[htb]
\centering
\resizebox{0.49\textwidth}{!}{%
\includegraphics{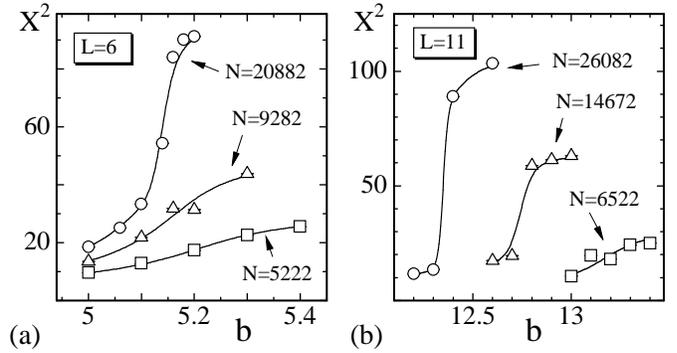}
}
\caption{The mean square size $X^2$ versus $b$ obtained on the surfaces of (a) $L\!=\!6$ and (b) $L\!=\!11$. The curves are drawn by the multihistogram reweighting technique.} 
\label{fig-11}
\end{figure}
Figures \ref{fig-11}(a) and \ref{fig-11}(b) are plots of $X^2$ against $b$ obtained under $L\!=\!6$ and  $L\!=\!11$. We find that the variation of $X$ becomes rapid against $b$ as $N$ increases. Thus, it is expected that the variation of $X^2$ has a jump at intermediate value of $b$ in either case of $L$. 

The bending energy $S_2^{(1)}/N_S^\prime$ is expected to reflect the transition, where $N_S^\prime$ is given by
\begin{equation}
\label{bendingenegryvertices}
N_S^\prime = N_S + 6N_J-12.
\end{equation}
$S_2^{(1)}/N_S^\prime$ is the bending energy per vertex, because $N_S^\prime$ is the total number of vertices where $S_2^{(1)}$ is defined. $N_S^\prime$ includes virtual vertices which are the corners of the junctions (see also Fig.\ref{fig-2}), which are not counted as vertices and hence are not included in $N_S$. Total number of virtual vertices are $6N_J-12$, because the hexagonal junction contains 6 virtual vertices, and the total number of pentagonal junction is $12$. Thus, we have Eq.(\ref{bendingenegryvertices}) for $N_S^\prime$, and therefore we have $N_S^\prime\!=\!1890$, $N_S^\prime\!=\!3360$, $N_S^\prime\!=\!5250$, and $N_S^\prime\!=\!7560$ for the length $L\!=\!6$ surfaces of size $(N,N_S,N_J)\!=\!(5222,1350,92)$, $(9282,2400,162)$, $(14502,3750,252)$, and $(20882,5400,362)$; and $N_S^\prime\!=\!1440$, $N_S^\prime\!=\!3240$, and $N_S^\prime\!=\!5760$ for the length $L\!=\!11$ surfaces of size $(N,N_S,N_J)\!=\!(6522,1200,42)$, $(14672,2700,92)$, and $(26082,4800,162)$.

\begin{figure}[htb]
\centering
\resizebox{0.49\textwidth}{!}{%
\includegraphics{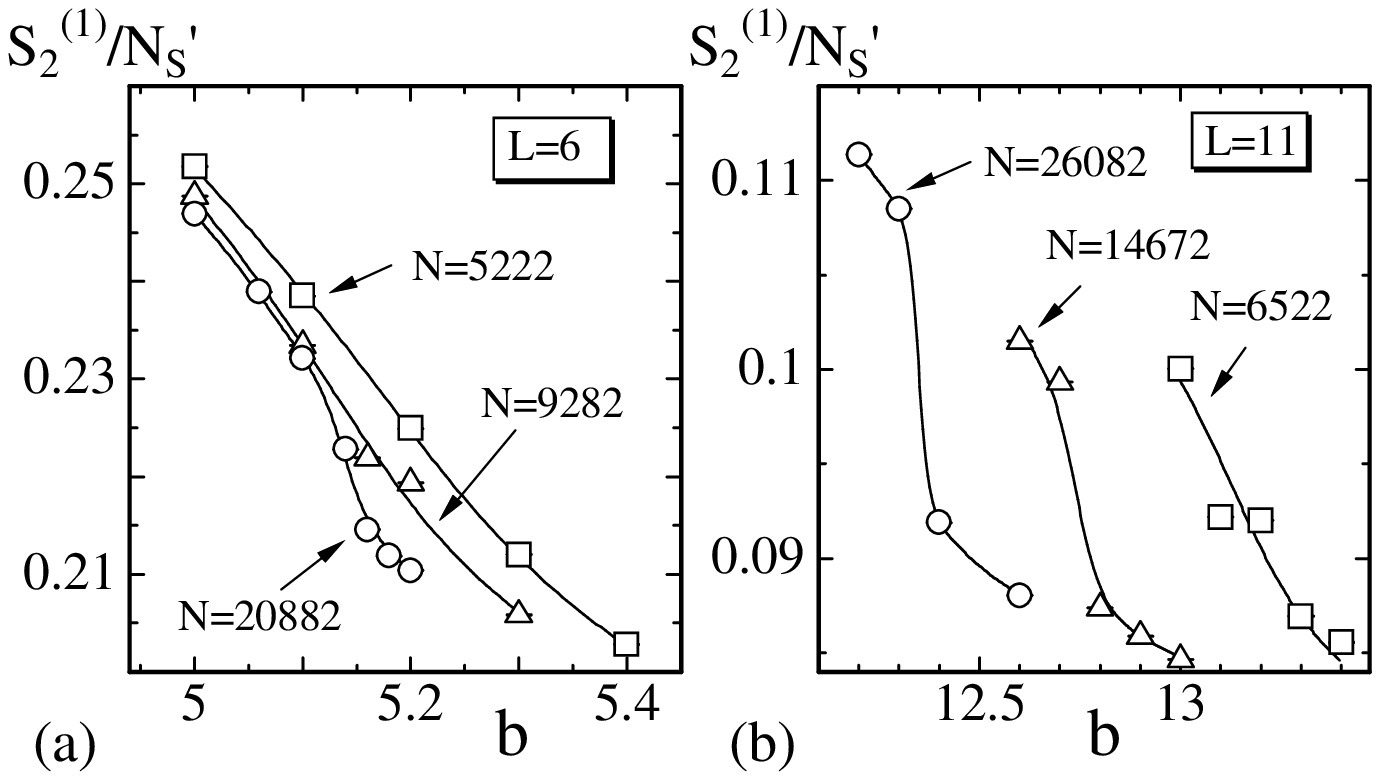}
}
\caption{The one-dimensional bending energy $S_2^{(1)}/N_S^\prime$ versus $b$ obtained on the surfaces of (a) $L\!=\!6$ and (b) $L\!=\!11$. } 
\label{fig-12}
\end{figure}

Figures \ref{fig-12}(a) and \ref{fig-12}(b) are plots of $S_2^{(1)}/N_S^\prime$ against $b$ obtained under $L\!=\!6$ and  $L\!=\!11$. We find the expected behavior in $S_2^{(1)}/N_S^\prime$ under both conditions  $L\!=\!6$ and  $L\!=\!11$; $S_2^{(1)}/N_S^\prime$ has a gap (or a jump) at intermediate $b$. This clearly shows that the surface fluctuation transition is of first order.

The transition can also be reflected in the two-dimensional extrinsic curvature, which is defined by
\begin{equation}
\label{two-dim-bending-energy}
 S_2^{(2)}\!=\!\sum_{\langle ij \rangle}\left(1-{\bf n}_i \cdot {\bf n}_j\right),
\end{equation}
where ${\bf n}_i$ is the unit normal vector of the triangle $i$. The definition Eq.(\ref{two-dim-bending-energy}) of $S_2^{(2)}$ is identical to that of $S_2$ in Eq.(\ref{Disc-Eneg-F}). In $S_2^{(2)}$, $\sum_{\langle ij \rangle}$ denotes the summation over all nearest neighbor triangles $i$ and $j$ that have the common bond $\langle ij \rangle$, which includes bonds belonging to the skeleton chains. $N_B$ denotes the total number of bonds where $S_2^{(2)}$ is defined, and it is given by $N_B\!=\! \sum_{\langle ij \rangle} 1$. Note that $N_B$ includes {\it virtual bonds}, which are the edges of the rigid junctions. In fact, we define $S_2^{(2)}$ even on the virtual bonds, because it is reasonable to define extrinsic curvature on those edges. It is also noted that $S_2^{(2)}$ is not included in the Hamiltonian, and therefore $S_2^{(2)}$ gives no mechanical strength to the surface.

\begin{figure}[htb]
\centering
\resizebox{0.49\textwidth}{!}{%
\includegraphics{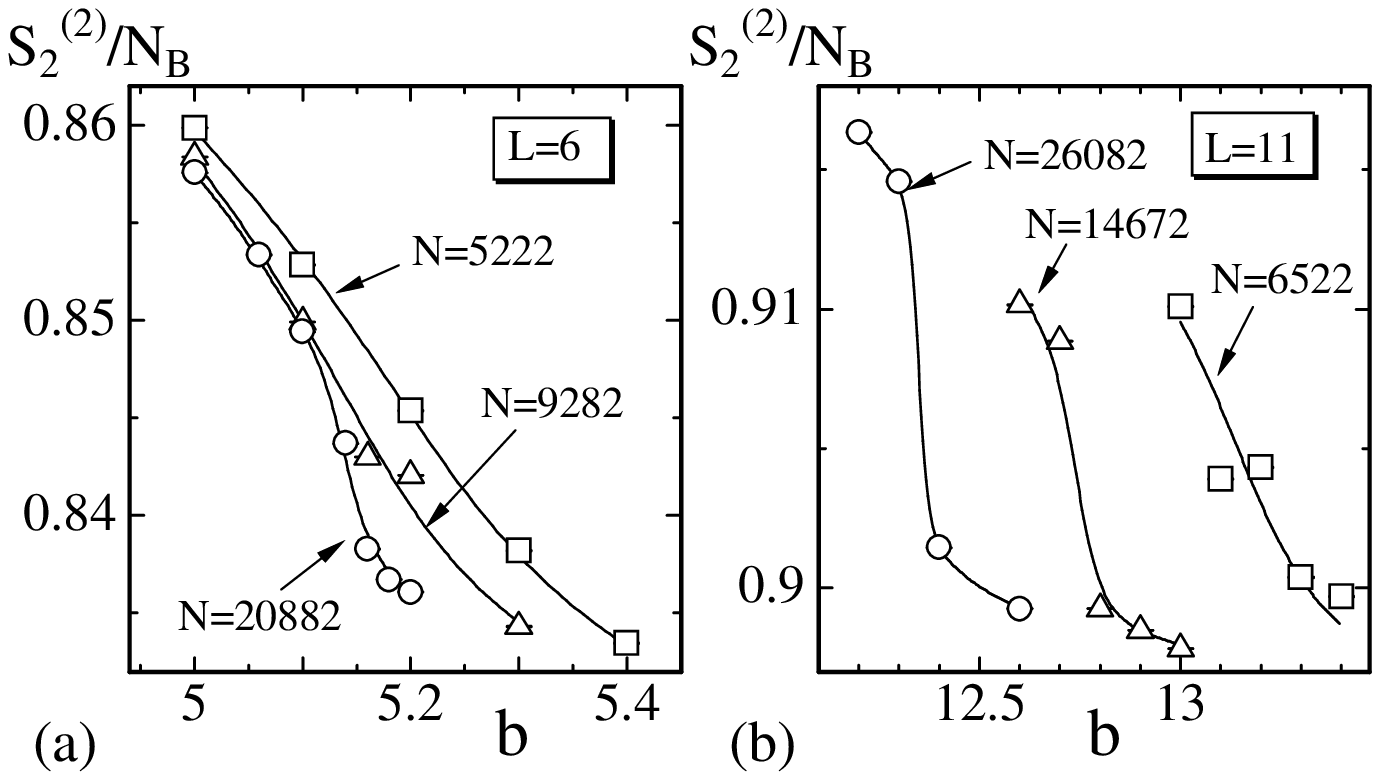}
}
\caption{The two-dimensional extrinsic curvature $S_2^{(2)}/N_B$ against $b$ obtained on the surfaces of (a) $L\!=\!6$ and (b) $L\!=\!11$. $S_2^{(2)}$ is defined by Eq.(\ref{two-dim-bending-energy}) and is not included in the Hamiltonian. $N_B$ is the total number of bonds where $S_2^{(2)}$ is defined.} 
\label{fig-13}
\end{figure}
The two-dimensional extrinsic curvature $S_2^{(2)}/N_B$ is plotted in Figs.\ref{fig-13}(a) and \ref{fig-13}(b) against $b$, which are corresponding to the lengths $L\!=\!6$ and  $L\!=\!11$. We find that the dependence of $S_2^{(2)}/N_B$ on $b$ shown in Fig.\ref{fig-13} is almost identical to that of $S_2^{(1)}/N_S^\prime$ in Fig.\ref{fig-12}. The gap (or jump) seen in  $S_2^{(2)}/N_B$ also supports that the surface fluctuation transition is of first order.

\begin{figure}[htb]
\centering
\resizebox{0.49\textwidth}{!}{%
\includegraphics{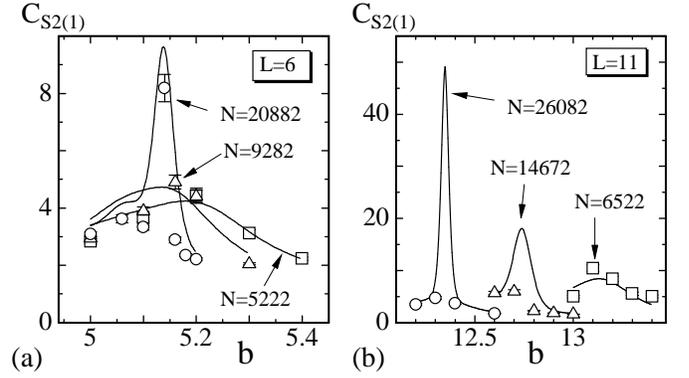}
}
\caption{The specific heat $C_{S_2^{(1)}}$ for $S_2^{(1)}$ versus $b$ obtained on the surfaces of (a) $L\!=\!6$ and (b) $L\!=\!11$. $C_{S_2^{(1)}}$ is defined by Eq.(\ref{specific-heat-1}). The error bars on the symbols are the statistical error, which is obtained by the binning analysis.} 
\label{fig-14}
\end{figure}

The specific heat corresponding to the one-dimensional bending energy $S_2^{(1)}$ is defined by 
\begin{equation}
\label{specific-heat-1}
C_{S_2^{(1)}} \!=\! {b^2\over N_S^\prime} \langle \; \left( S_2^{(1)} \!-\! \langle S_2^{(1)} \rangle\right)^2\rangle,
\end{equation}
which can also reflect phase transitions if it has an anomalous behavior. Figures \ref{fig-14}(a) and \ref{fig-14}(b) show $C_{S_2^{(1)}}$ versus $b$ obtained under $L\!=\!6$ and  $L\!=\!11$. Solid curves drawn in the figures were obtained by the multihistogram reweighting technique, and those curves apparently show an expected anomalous behavior indicating that $C_{S_2^{(1)}}$ is divergent when $N_S^\prime\to\infty$ (or equivalently $N\to\infty$).   

\begin{figure}[htb]
\centering
\resizebox{0.49\textwidth}{!}{%
\includegraphics{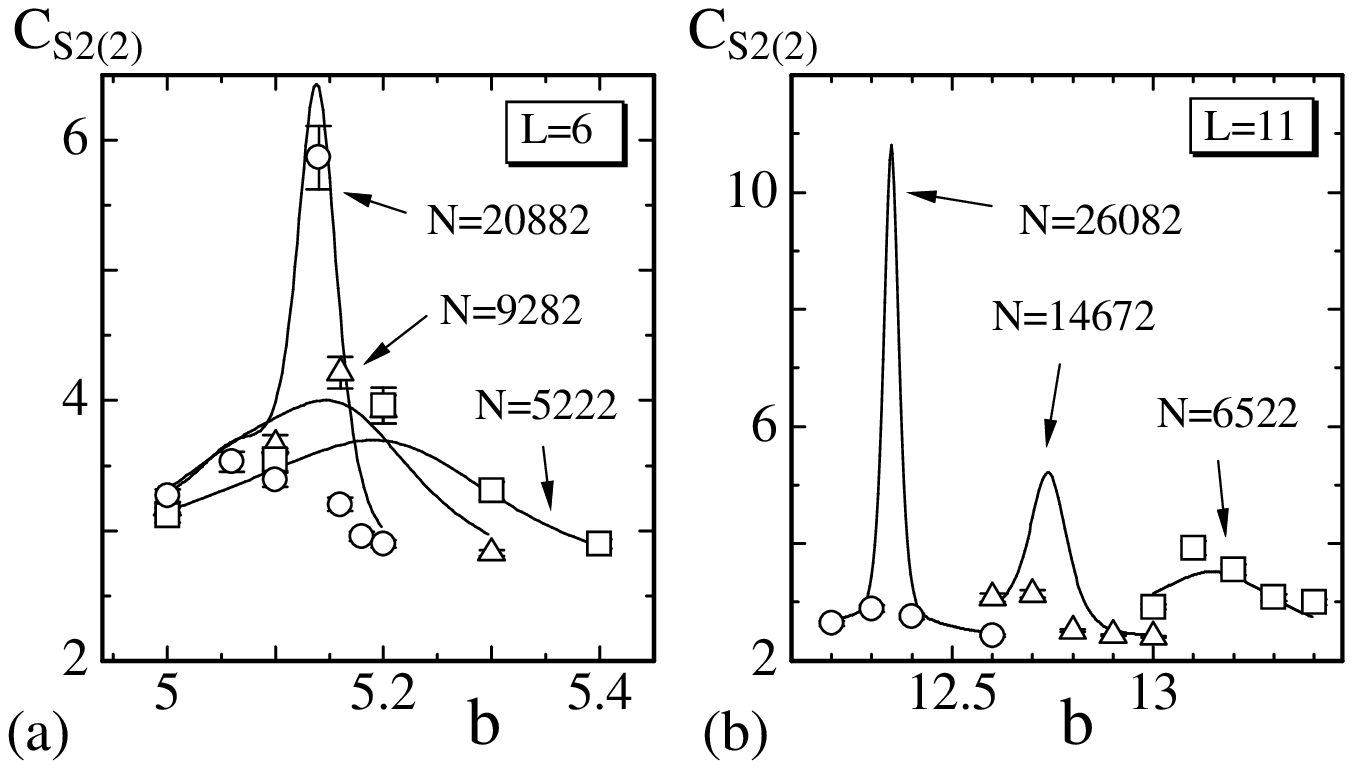}
}
\caption{The specific heat $C_{S_2^{(2)}}$ for $S_2^{(2)}$ versus $b$ obtained on the surfaces of (a) $L\!=\!6$ and (b) $L\!=\!11$. $C_{S_2^{(2)}}$ is defined by Eq.(\ref{specific-heat-2}). The error bars on the symbols are the statistical error, which is obtained also by the binning analysis.} 
\label{fig-15}
\end{figure}
The specific heat corresponding to the extrinsic curvature $S_2^{(2)}$ in Eq.(\ref{two-dim-bending-energy}) can also be defined by 
\begin{equation}
\label{specific-heat-2}
 C_{S_2^{(2)}} \!=\! {1 \over N} \langle \; \left( S_2^{(2)} \!-\! \langle S_2^{(2)} \rangle \right)^2 \rangle, 
\end{equation}
which reflects the transition as $C_{S_2^{(1)}}$ does. Curvature coefficient for $C_{S_2^{(2)}}$ was assumed to be $1$, because $S_2^{(2)}$ is not included in the Hamiltonian.  Figures \ref{fig-15}(a) and \ref{fig-15}(b) show $C_{S_2^{(2)}}$ against $b$ obtained under $L\!=\!6$ and  $L\!=\!11$. We can see in $C_{S_2^{(2)}}$ the same anomalous behavior as in $C_{S_2^{(1)}}$. 

\begin{figure}[htb]
\centering
\resizebox{0.49\textwidth}{!}{%
\includegraphics{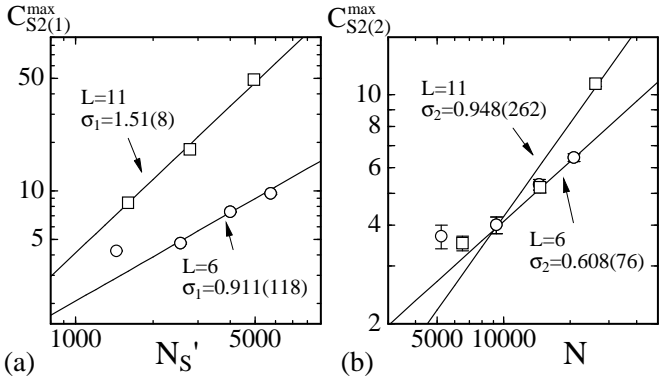}
}
\caption{ Log-log plots of (a) $C_{S_2^{(1)}}^{\rm max}$ against $N_S^\prime$ and (b) $C_{S_2^{(2)}}^{\rm max}$ against $N^\prime$ obtained on the surfaces of $L\!=\!6$ and $L\!=\!8$. The straight lines are drawn by fitting the largest three data of $C_{S_2^{(1)}}^{\rm max}$ and $C_{S_2^{(2)}}^{\rm max}$ to Eq.(\ref{scaling-exponents}). The peak values and the statistical errors for the fittings were obtained by the multihistogram reweighting.} 
\label{fig-16}
\end{figure}
In order to see the anomalous behaviors in $C_{S_2^{(1)}}$ and $C_{S_2^{(2)}}$ in more detail, we plot the peak values of them in Figs.\ref{fig-16}(a) and \ref{fig-16}(b) in log-log scales against $N_S^\prime$ and $N$, respectively. The straight lines were drawn by fitting the data to 
\begin{equation}
\label{scaling-exponents}
C_{S_2^{(1)}}^{\rm max} \propto \left( N_S^{\prime}\right)^{\sigma_1}, \quad C_{S_2^{(2)}}^{\rm max} \propto \left( N \right)^{\sigma_2}, \quad 
\end{equation}
where $\sigma_1$, $\sigma_2$ are critical exponents. Largest three data were contained in the fitting in the case $L\!=\!6$. Thus, we have
\begin{eqnarray}
\label{exponents-values}
\sigma_1=0.911\pm 0.118, \; \sigma_2=0.608\pm 0.076, \; (L=6), \nonumber \\
\sigma_1=1.51\pm 0.08, \; \sigma_2=0.948\pm 0.262, \; (L=11).   
\end{eqnarray}
The result $\sigma_2\!=\!0.608(76)$ for $L\!=\!6$ is inconsistent to the fact that the surface fluctuation transition is of first-order, however, $\sigma_1\!=\!0.911(118)$ is consistent to the discontinuous collapsing transition. The results $\sigma_1\!=\!1.51(0.08)$ and $\sigma_2\!=\!0.948(262)$ under $L\!=\!11$ support the discontinuous transition of surface fluctuation.

\section{Summary and Conclusion}
We have studied a compartmentalized fluid surface model and found that the model undergoes a first-order collapsing transition and a first-order surface fluctuation transition between the smooth phase and the collapsed phase. The model is classified as a fluid surface model, although the long-range order and the phase transition can be seen at finite $b$. The compartmentalized structure is considered to be a reason for the existence of the phase transition. Consequently, the result is not in contradiction with the standard argument for the non-existence of long-range order in fluid membranes.  Moreover, the critical point of the transition is strongly expected at finite $n_c(\simeq 91)$, where the collapsing transition and the surface fluctuation transition terminate and turn into continuous or higher-order transitions. In fact, we demonstrated that the collapsing transition remains first-order at $n\!\leq\!91$ and disappear at $n\!=\!120$ on sufficiently large surfaces. It was also demonstrated that the phase transition of surface fluctuation remains first-order at $n\!\leq\!66$ and disappear at $n\!=\!120$ on sufficiently large surfaces, and that the transition weakens and still survives at $n\!=\!91$. 

A surface model with skeletons has also been investigated by using the canonical Monte Carlo simulations. The skeletons are composed of one-dimensional linear chains and rigid junctions, whose size is chosen sufficiently small compared to the mean bond length. The surface is a triangulated sphere and divided into a lot of compartmentalized domains, whose boundary corresponds to the skeletons, and it is almost identical to that for the compartmentalized fluid model. The skeleton surface is characterized by $(N,N_S,N_J,L)$, which are respectively the total number of vertices including the junctions, the total number of vertices on the chains, the total number of junctions, and the length of chains between junctions. The length of chains was fixed to $L\!=\!6$ and $L\!=\!11$, which correspond to $n\!=\!21$ and  $n\!=\!66$ the total number of vertices in a compartment.   

The mechanical strength is given to the surface only by the skeletons. There is no two-dimensional curvature energy in the Hamiltonian, while one-dimensional bending energy is defined on the compartment boundary. The two-dimensional Gaussian bond potential is included in the Hamiltonian just like in the standard surface model of Helfrich, Polyakov and Kleinert.  The skeleton model in this paper is different from the one with elastic junctions in \cite{KOIB-JSTP-2007-1}, because the rigid junctions cannot be identified with the elastic junctions due to the property on the in-plane elasticity at the junctions. 

We found that the skeleton surface in this paper undergoes a first-order collapsing transition and a first-order surface fluctuation transition between the smooth phase and the crumpled phase. The one-dimensional bending energy $S_2^{(1)}$ has a gap (or a jump) at intermediate bending rigidity $b$, and the two-dimensional extrinsic curvature $S_2^{(2)}$also has a gap at that point. These imply that the surface fluctuations are considered to be a first-order transition. Moreover, it is found that the mean square size $X^2$ also has a gap at the transition point. This implies that the surface-collapsing phenomenon can be viewed as a first-order transition. 

The results in this paper together with those in \cite{KOIB-JSTP-2007-1} show that the first-order transitions can be seen in the spherical surface model even when the mechanical strength is maintained only by skeletons, which are composed of linear chains joined to each other at the junctions. Moreover, the order of transitions is independent of whether the junction is elastic or rigid. 

We have studied two types of compartmentalized surface models; the mechanical strength is maintained by the two-dimensional curvature in the first model (the fluid model), and it is maintained by the one-dimensional curvature in the second model (the skeleton model). Therefore, we can also conclude that the first-order transitions occur independent of whether the shape of surface is mechanically maintained by the skeleton (= the domain boundary) or by the surface itself.

Vertices can freely diffuse inside each compartment on the fluid surfaces supported by the skeletons. It is interesting to see how fluidity influences the transition of the skeleton-supported model. Many interesting problems remain to be studied on the surface model with skeletons.

This work was supported in part by Grant-in-Aid for Scientific Research, No. 15560160 and No. 18560185.  

\end{document}